\documentclass[aip,jcp,preprint]{revtex4-1}

\usepackage{graphicx,graphics,color}
\usepackage{amssymb, amsmath,array,float}
\usepackage[utf8]{inputenc}

\usepackage{graphicx}
\usepackage{dcolumn}
\usepackage{bm}

\newcommand{\dint}{\displaystyle \int}

\begin{document}

\title{Full quantum dynamics of the electronic coupling between photosynthetic pigments}

\author{M. Bel\'en Oviedo}
\altaffiliation[Present Address: ]{Department of Chemical \& Environmental Engineering and Materials Science, University of California, Riverside, California 92521, United States}
\author{Cristi\'an G. S\'anchez}
\email{cgsanchez@fcq.unc.edu.ar}
\affiliation{Departamento de Matem\'atica y F\'{\i}sica, Facultad de Ciencias Qu\'{\i}micas, INFIQC, Universidad Nacional de C\'ordoba, Ciudad Universitaria, X5000HUA, C\'ordoba, Argentina}

\begin{abstract}
From studying the time evolution of the single electron density matrix within a density functional tight-binding formalism we study in a fully atomistic picture the electronic excitation transfer between two photosynthetic pigments in real time. This time-dependent quantum dynamics is based on fully atomistic structural models of the photosynthetic pigment. We analyze the dependence of the electronic excitation transfer with distance and orientation between photosynthetic pigments. We compare the results obtained from full quantum dynamics with analytical ones, based on a two level system model were the interaction between the pigments is dipolar. We observed that even when the distance of the photosynthetic pigment is about $30$ \AA{} the deviation of the dipolarity is of about $15$ percent.
\end{abstract}


\maketitle

\section{Introduction}
Photosynthesis is a natural process that begins with the absorption of sunlight by an arrangements of photosynthetic pigments embedded into a proteic matrix known as light harvesting (LH) antenna complexes. In this process, the electronic excitation (known as exciton) spreads and moves within and between pigments within the complex, eventualy reaching the reaction center where charge separation occurs \cite{Blakenship}, leading to the biochemical energy conversion process. The quantum efficiency of the early event, namely, the absorption of a photon and energy transfer, is very close to $100$\%. A number of studies on photosynthetic complexes suggests that this high efficiency can be due to suppression of environmental decoherence of excitons within antenna complexes \cite{EngelCRAMCBF2007, Fleming2007,Calhoun2009Fleming} even a room temperature \cite{Collini2010Scholes,Engel2010,Harel2012Engel,Fleming2012}. Protein matrices in the antenna complex appear to suppress decoherence processes and allow the joint excitation of many photosynthetic pigments and the consistent evolution of the exciton for a relatively long time similar to the electronic excitation transfer (EET) time scales, allowing the parallel exploration of a manifold of relaxation paths during this period \cite{WangLAWBLW2007,Aspuru-Guzik2008,Panitchayangkoon2011MukamelEngel}. This evidence implies that quantum coherences should be taken into account during the study of energy transfer in photosynthetic antenna systems.

Understanding the physical principle underlying highly efficient energy transfer in photosynthesis has motivated researchers to develop several theoretical models with this aim. EET within photosynthetic complexes is usually described by two perturbative limits depending on the magnitude of electron-phonon coupling. When the electronic coupling between the photosynthetic pigments is small, this coupling is described in terms of inter-pigment dipole-dipole interaction which is the lowest order nonvanishing term in the multipole expansion. The two assumptions underlying this approach are that vibrational relaxation within the excited state is much faster than energy transfer, and that the coupling to the vibrational modes of the bath is stronger than the coupling between the chromophores. The latter approach ensures that EET occurs via an incoherent hopping mechanism and shows Markovian behavior \cite{mukamel,may2011charge}. In this regime the rate of EET is generally described by F\"orster theory \cite{scholes2005fleming}, where this rate can be expressed in terms of the spectral overlap of the donor emission spectrum and the acceptor absorption spectrum. This approach has been implemented for describing the physics involved in the study of EET in the coupled B800 dimer ring of the purple bacteria LH2 complex \cite{Sundstrom1999PGrondelle,Hofmann2003KMMAK}.

Furthermore, in the strong electronic coupling case, the excited states of the pigments mix and new delocalized states emerge. On short time scales, this delocalization leads to coherent non-Markovian dynamics in which the excitation travels as a wave until dephasing destroys the coherence \cite{EngelCRAMCBF2007,Cheng2009Fleming}. When the distance between the chromophores is similar to the spatial extent of the molecules, electronic wavefunctions begin to overlap and electronic exchange becomes important. Dexter theory includes the exchange interaction in the description of the EET \cite{Burghardt2009}. However, in the case where the electron-phonon coupling is strong, neither F\"orster nor Dexter are sufficient to describe the EET. Where EET can occur on timescales faster than vibrational relaxation \cite{Ishizaki2009,Ishizaki2009Fleming} Redfield or Lindblad relaxation theory are used \cite{Ishizaki2009,Ishizaki2009Fleming} in order to include explicitly the coupling between chromophores and some thermal bath vibrational modes.

Theoretical approaches applied to study this phenomenon are based on parametrized Hamiltonians containing information about inter-chromophore couplings and the excitation energy of individual pigments embedded in their proteic environment. Most of these parameters are obtained from experimental data \cite{Schulten2011,scholes2005fleming,Ishizaki2009Fleming,Adolphs2006Renger,Olbrich2010,Hsin2010Shulten}. Furthermore, the description of the inter-chromophore coupling can be obtained from quantum chemical approaches like transition density cube (TDC), a three dimensional (3D) {\em ab initio} approach for the calculation of the EET couplings. This method gives a more realistic representation of the transition density of the molecules, since it takes into account the shape of the interacting molecules \cite{krueger1998SholesFleming}. Madjet {\em et al.} \cite{Madjet2006} presented a numerically more efficient method for the calculation of the inter-chromophore coupling, known as TrEsp (transition charges from electrostatic potential method). The coupling is obtained by fitting the electrostatic potential of the transition density on a 3D grid using atomic partial charges. 

Despite the existence of a variety of models and theoretical methods that complement the experimental results, to our knowledge, there is currently no method that can describe the photophysics of individual photosynthetic pigments as well as the dynamical evolution of their coupled excitations when embedded within a proteic environment from an atomistic, time dependent perspective \cite{Buda2009,Scholes2011}. The main problem lies in that due to the large size of the antenna complex, i.e. a few thousand atoms, this entire system can not be simulated with the present computational methods found in literature \cite{Buda2009,Renger2009}.

In a previous work we calculate the excitation energies as well as the transition dipole moments vector for a series of important photosynthetic pigments \cite{Oviedo2010, Oviedo2011}. In this paper we go a step further towards the comprehension of the EET mechanism. We study the full dynamical evolution of the one electron density matrix within a density functional tight-binding (DFTB) Hamiltonian in response to laser illumination. From this simulations we obtain the expectation value of the dipole moment, which takes into account the entire structural information of the molecule. From linear response theory, considering each pigment as a two level system (TLS) we obtain the analytical expressions of this expectations values and compare them with the calculated ones. As expected, the interaction between a chlorophyll $a$ (Chla) dimer deviates from the dipole-dipole interaction when the distance between them is less or similar to the spatial extent of the wave function of each pigment, however when the dimer is about $30$ \AA{} the interaction deviates by a 10\%. This is not a minor result given that the models used to simulate the couplings between the photosynthetic pigments in the study of the energy transfer takes into account only dipolar interactions \cite{Mohseni2008Alan,Schulten2011}. We describe as well deviations from the dipolar coupling model as a function of the orientation of dimer components.

\section{Computational Method}

The description of the electronic structure of photosynthetic pigments was carried out using the self consistent density functional tight-binding (SCC-DFTB) method\cite{Porezag1995SeifertFrauenheim,Elstner1998PJEHFrauenheimSS}. This method has been successfully applied to the description of the electronic structure of large molecular systems\cite{Elstner1998PJEHFrauenheimSS}. It is based on a second order expansion of the Khon-Sham energy functional around a reference density of neutral atomic species. The DFTB+ code is computational implementation of the DFTB method. We have used this code to obtain the Hamiltonian, the overlap matrix and the ground state single electron density matrix. Our implementation differs from that of reference \cite{Niehaus2005HTFrauenheim} in that it propagates the one electron density matrix instead of the single particle orbitals. We have successfully applied this model to the calculation of chlorophyll spectra, transition dipole moment direction and magnitude. We obtained excellent agreement with experimental and theoretical results \cite{Oviedo2010,Oviedo2011}. For the geometries of the photosynthetic pigments used in \cite{Oviedo2010,Oviedo2011} we applied to their ground state density matrix a classical sinusoidal electric field in the dipole approximation:
\begin{equation}\label{eq-field}
\hat{H}=\hat{H}_0 + \mathbf{E_0} \sin(\omega_{Q_y}t)\cdot\hat{\boldsymbol\mu}
\end{equation}
where the field is applied in the direction and in tune with the $Q_y$ electronic transition \cite{Oviedo2011}. The evolution of the system can be calculated by integrating the Liouville-Von Newmann equation of motion in the non-orthogonal basis:
\begin{equation}
\frac{\partial \hat{\rho}}{\partial t}=\frac{1}{i\hbar}\left(S^{-1}\hat{H}[\hat{\rho}]\hat{\rho}-\hat{\rho}\hat{H}[\hat{\rho}]S^{-1}\right)
\label{rhodot}
\end{equation}

The dipole moment is calculated as:
\begin{equation}\label{mu-dftb}\displaystyle
 \boldsymbol{\mu}(t)=e\sum_i^N q_i(t) \mathbf{r}_i 
\end{equation}
where $e$ is the elementary charge, $\mathbf{r}_i$ is the cartesian coordinate of atom $i$ and $q_i(t)$ is the Mulliken charge given by
\begin{equation}\displaystyle
 q_i(t)=\sum_{\nu\in i}^N Z_{\nu}-(\rho S+S\rho)_{\nu\nu}
\end{equation}
were $Z$ is effective nuclear charge associated to atom $i$.

To study the interaction between the chromophores within the dipole approximation, we consider each photosynthetic pigment as a two level system (TLS), where the difference between each energy level is equal to the $Q_y$ electronic transition. In the linear regime, where the intensity of the electric field is small, the response of the dipole moment, in matrix notation, to the laser perturbation is:
\begin{equation}\label{eq-rta}
  \mu_{\alpha}(t)=-\dfrac{i}{\hbar} \dint^{\infty}_{0} d\tau \sum_{\beta} \langle[\hat{\mu}_{\alpha}(\tau),\hat{\mu}_{\beta}]\rangle E_{\beta}(t-\tau)
\end{equation}
where $\alpha$ equals to $x$, $y$ or $z$ and $\langle[\hat{\mu}_{\alpha}(\tau),\hat{\mu}_{\beta}]\rangle$ is the polarizability tensor that describes the dipole moment response in direction $\alpha$ to an applied electric field in direction $\beta$. The expression of this tensor for a TLS is given by the following expression:
\begin{equation}
  \langle[\hat{\mu}_{\alpha}(\tau),\hat{\mu}_{\beta}]\rangle=\dfrac{2i}{\hbar} |\mu_{Q_y}|^2\sin\left(\tau \omega_{Q_y}\right)r^{\alpha}_{Q_y}r^{\beta}_{Q_y}
\end{equation}
where $|\mu_{Q_y}|^2$ is the transition dipole moment, $\omega_{Q_y}= \frac{\Delta  E_{Q_y}}{\hbar}$ and $\hat{r}_{Q_y}=r^x_{Q_y}\hat{i}+r^y_{Q_y}\hat{j}+r^z_{Q_y}\hat{k}$ such that the inner product of $\hat{r}_{Q_y}$ is equal to $1$. 

If the applied field is in the direction of the $Q_y$ electronic transition, then the expectation value of the dipole moment is equal to:
\begin{equation}\label{eq7}
  \mu_{\alpha}(t)=\dfrac{2}{\hbar} E_0|\mu_{Q_y}|^2\dint^{\infty}_{0} d\tau \sin(\omega_{Q_y}\tau)\sin(\omega_{Q_y}(t-\tau))r^{\alpha}_{Q_y}
\end{equation}
The solution of eq. (\ref{eq7}) for long times can be approximated by the following:
\begin{equation}
  \boldsymbol{\mu}(t)\simeq \dfrac{E_0}{\hbar}|\mu_{Q_y}|^2\cos\left(t\omega_{Q_y}\right)t\:\hat{r}_{Q_y}
\label{muD}
\end{equation}

\begin{figure}\includegraphics[width=6cm]{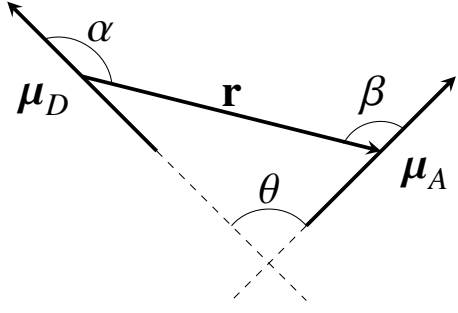}
\caption{Schematic representation of two Chl's molecules separated by the distance $\mathbf{r}$. The molecules are characterized by their dipole moment, $\boldsymbol{\mu^I}$, $\theta$ is the angle between the two dipoles moments, $\alpha$ and $\beta$ are the angles between each dipole moment and the distance vector, $\mathbf{r}$.} \label{fig1}
\end{figure}

Suppose that the second chromophore ({\bf A}) is located at $\vec{r}$ from the first one ({\bf D}) and $|\vec{r}|$ is larger than the spatial extent of {\bf A} and {\bf D}, the electric field generated by the oscillating dipole of {\bf D} is given by the following equation
\begin{equation}\label{ef}
 \mathbf{E}_D(t)=\frac{1}{4\pi\epsilon_0r^3}\left(3(\boldsymbol{\mu}_D(t)\cdot\mathbf{\hat{r}})\cdot\mathbf{\hat{r}}-\boldsymbol{\mu}_D(t)\right)
\end{equation}
where $\boldsymbol{\mu}_D(t)$ is the expectation value of the dipole moment of {\bf D} given by eq. (\ref{muD}). $\epsilon_0$ is the vacuum permittivity and $r$ is the distance between the Mg atoms (see Fig.~\ref{fig1}).

We assumed that an electric field given by eq. (\ref{ef}) perturbs the second molecule, substituting the expression of the electric field into eq. (\ref{eq-rta}), the expectation value of the dipole moment of molecule {\bf A} is:
\begin{widetext}
\begin{equation}\label{muA-ang}
 \boldsymbol{\mu}_A(t)\simeq \dfrac{E_0}{4\pi\epsilon_0\hbar^2r^3}|\mu_{D}|^2|\mu_{A}|^2\sin\left(t\omega_{Q_y}\right)t^2\left(\cos(\beta)\cos(\alpha)-\dfrac{1}{2}\sin(\beta)\sin(\alpha)\right)\:\hat{r}^A_{Q_y}
\end{equation}
\end{widetext}
where $|\boldsymbol{\mu}_A|^2$ and $|\boldsymbol{\mu}_D|^2$ are the transition dipole moment of the acceptor and donor molecules, respectively. $\hat{r}^A_{Q_y}$ is the direction of the $Q_y$ electronic transition of the acceptor molecule,
$\alpha$ and $\beta$ are the angles between each dipole moment and the distance vector, $\mathbf{r}$ (Fig.~\ref{fig1}).

\section{Results and Discussion}
\begin{figure}\includegraphics{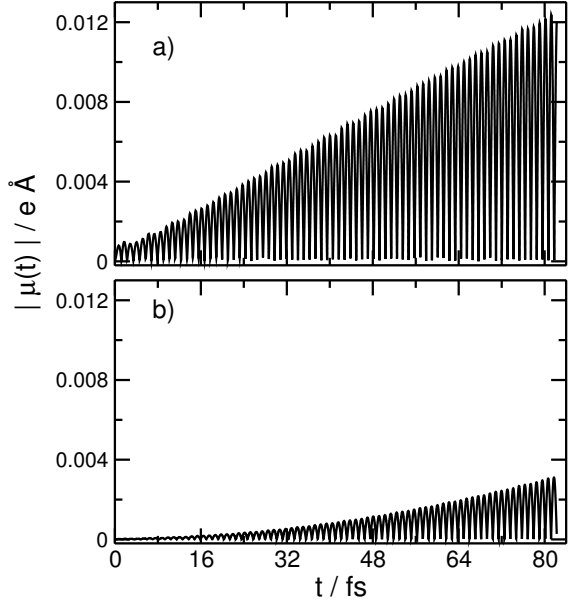}
\caption{Absolute value of the dipole moment recorded as a function of time of individual molecules of Chl $a$ in response to laser illumination within the time-dependent DFT description. The external field is applied to one of the molecules (a) while the other (b) is stimulated by the full oscillating field of the first, including direct electronic couplings given by matrix elements within the DFTB Hamiltonian. The distance between Mg atoms is $21.0$ \AA{} and the molecules are aligned along the $Q_y$ electronic transition.} \label{fig2}
\end{figure}

As a first step towards the understanding of the interaction nature between the photosynthetic pigments, we begin analyzing the coupling between two Chla molecules from a full quantum dynamical simulation without any approximations other than those implied in the nature of the DFTB Hamiltonian. We studied the variation of the interaction when distance and orientation between the molecules change independently. In both instances we perform a comparison between the dynamical results and the analytical ones and we show the percentage of deviation from the ideal dipole-dipole interaction.

The quantum dynamical simulation starts with the perturbation of one of the Chla's with a laser type perturbation (\ref{eq-field}) in tune with the electronic excitation ($\omega_{Q_y}=1.91$ eV) and in the direction of the $Q_y$ transition. The intensity of the perturbation was of $E_0=0.1$ mV/\AA{} in order to remain in the linear response regime. The molecules are aligned in the $Q_y$ direction and the system was allowed to evolve in time, then we calculated the dipole moment of each chromophore (equation \ref{mu-dftb}). Fig.~\ref{fig2} describes the EET from the Chla wich is illuminated (donor) to the second molecule (acceptor). Fig.~\ref{fig2} (a) shows the time dependent variation of the dipole moment of the monomer at which the external field is applied. It can be observed that the dipole moment oscillation grows linearly with time, as can be expected for an applied field within the linear response regime and in the absence of any dissipative mechanism. Fig.~\ref{fig2} (b) is the time evolution of the dipole moment of the second molecule in response to the stimulation by the oscillating field generated by the first. It is worth noting that besides the electrostatic coupling molecules may interact via non-zero Hamiltonian matrix elements within the DFTB Hamiltonian. 
This results are in agreement with the analytical expressions, eq. \ref{muD} and \ref{muA-ang} (with $\alpha$ and $\beta$ zero) derived in the previous section. As can be seen the oscillation of the dipole moment of the non-directly illuminated monomer grows quadratically in time, and depends on the distance to the first molecule as $r^{-3}$.

\begin{figure}\includegraphics{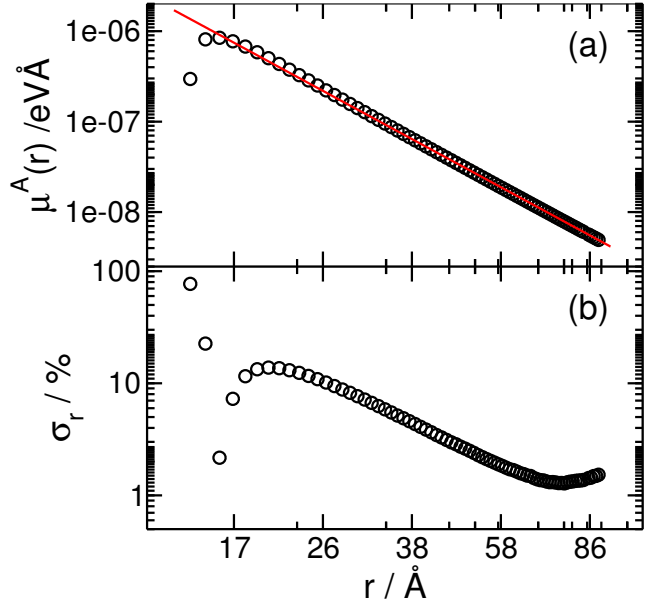}
\caption{(a) Log-log plot of the dipole moment of the acceptor molecule $\mu_A$ and (b) the percentual relative error ($\sigma_r$) as function of the distance between corresponding magnesium atoms. The solid line corresponds to the analytical function of the dipole moment using as transition dipole moment for Chl $a$ $1.03$ e$^2$ \AA$^2$ \cite{Oviedo2011}.} \label{fig3}
\end{figure}
The oscillation amplitudes of the dipole moment corresponding to the acceptor molecule (where no laser perturbation is applied) versus the distance between the magnesium atoms, $r$, are plotted in Fig.~\ref{fig3} (a). The solid line is the plot of the analytical function of the dipole moment given by eq. (\ref{muA-ang}). Where we have used as value of the transition dipole moment of the molecule D and A the value obtained for the monomeric photosynthetic pigment, this is $1.03$ e$^2$\AA{}$^2$ \cite{Oviedo2011}. As can be noted from the plot, for long disntances the EET can be described through the dipolar mechanism. However, when the distance between the molecules is around $40$ \AA, the transfer of the excitation starts to deviate from the dipolar mechanism, and a more detailed description of the electronic coupling between pairs of pigments should be considered, where it must be taken into account higher order terms in the multipole expansion and the description of the overlap between molecular orbitals.

Furthermore, it three regimes can be distinguished for the deviation from the dipolar mechanism: at long distances ($r>60$ \AA), the slope of the line corresponding to the analytical function is lower than the line product of the oscillations amplitudes of the dipole moment calculated by the electron dynamics, this observation indicates an error by excess and can be attributed to a systematic error caused by the fitting of the oscillations amplitudes of the dipole moment with a quadratic function. This error increases with the distance between the photosynthetic pigments, since the magnitude of the dipole of the acceptor molecule decreases with $r^{-3}$. The second regime is observed at intermediate distances ($17$ \AA{} $<r<60$\AA), where the error is by excess and is due to the absence of higher order terms in the multipolar expansion that describes the interaction between the molecules. Finally an abrupt change in the regime can be observed, where the error is by defect and is attributed to no taking into account the overlap between the electronic states of the two molecules in the multipolar expansion. This overlap allows direct charge transfer between chromophores, disabling EET due to the decrease of the excitation cross-section of the chromophores ($|\mu_{Q_y}|$), which implies a decrease in the intensity of the oscillating electric field ($\mathbf{E}_D(t)$) around the acceptor molecule.

\begin{figure}\includegraphics{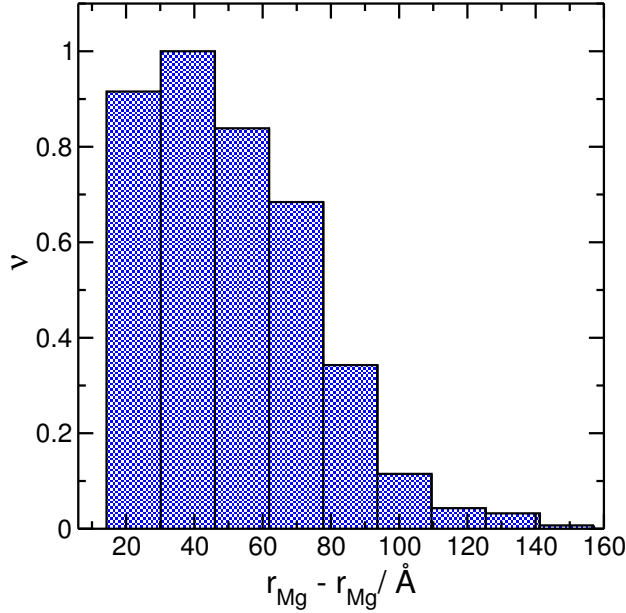}
\caption{Histogram of the distance between photosynthetic pigments in the complex antennas most common in nature.\cite{FMO,Peridinin,PSI,PSII,LH1,LH2}}  \label{fig4}
\end{figure}
This is not a minor result since the computational methods used to model the electronic coupling between the photosynthetic pigments in order to study the the energy transfer takes into account only the dipolar term \cite{Aspuru-Guzik2008, Schulten2011}. In photosynthetic organisms the is a remarkable variety of antennas, which show no apparent relation to each other in terms of structure or even types of pigments utilized \cite{Blakenship}. We constructed a histogram of the distance between the Mg atoms of all dimers of photosynthetic pigments present in antenna complexes found in nature for which the crystal structure has been described \cite{FMO,Peridinin,PSI,PSII,LH1,LH2} (Fig.~\ref{fig4}). It can be seen that most of the pigments are at $30$ \AA\ from each other. Therefore, if the dipole-dipole interaction is used to describe the interaction between the photosynthetic pigments, the results will show an error of around $15$ \% for these dimers (Fig.~\ref{fig3}(b)). This result shows that the use the dipole-dipole interaction approximation to describe their coupling is not appropriate at relevant interpigment distances, due both to the the influence of non-dipolar terms and direct electronic coupling between pigments.

\begin{figure}\includegraphics[width=7cm]{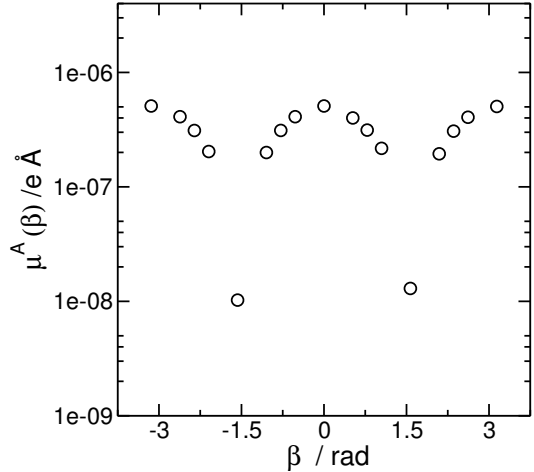}
\caption{Dependence of the dipole moment of the acceptor molecule, $\boldsymbol{\mu_A}$ with the orientation of this dipole moment with respect $\beta$. the distance between the two Mg atoms of the chromophores is $20$ \AA{}.} \label{fig5}
\end{figure}
In a second step we study the dependence of the electronic excitation transfer with the orientation of the molecules (equation \ref{muA-ang}).
For this purpose a continuous laser was applied to one Chla in tune with the $Q_y$ electronic excitation and the dipole moment of the acceptor Chla located at $20$ \AA{} was computed. Several simulations were performed by varying the orientation of the acceptor molecule, and Fig.~\ref{fig5} shows the orientation dependence for the dipole moment of thhis molecule. It can be seen that this variation has the shape of a cosine function. When the two pigments are oriented perpendicular to each other ($\beta=\pi/2$ rad) the dipole moment is zero, indicating that the second molecule is not excited and the maximum excitation is reached when the two pigments are aligned with the $Q_y$ electronic transition ($\beta=0$ rad and $\beta=\pi$ rad). 

\begin{figure*}\includegraphics{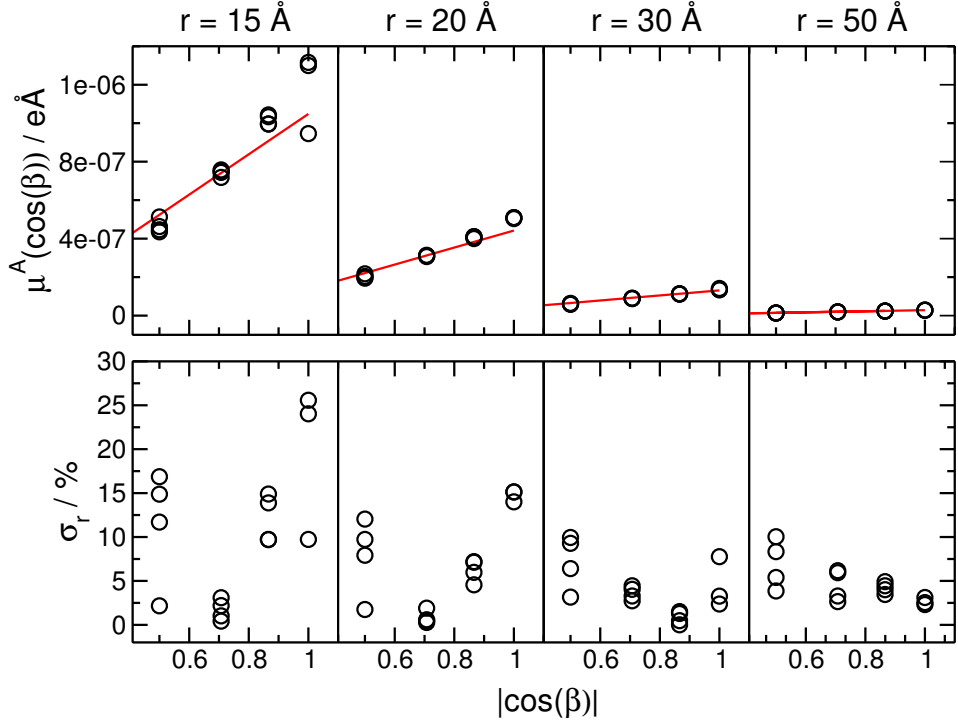}
\caption{Dependence of the dipole moment of the acceptor molecule (upper panel) and the percentage of the relative error (lower panel) with $|\cos(\beta)|$.}  \label{fig6}
\end{figure*}

To analyze in more detail the deviation from dipolar behavior, Fig.~\ref{fig6} shows the dependence of the dipole moment of the acceptor molecule with the $|\cos(\beta)|$ for different distances between the Mg atoms (upper panel) and the percentage relative error between the analytical results (eq. (\ref{muA-ang})) and the results obtained from eq. (\ref{mu-dftb}) (lower panel). 
If the interaction is dipolar the dipole moment of the acceptor molecule varies linearly with respect $|\cos(\beta)|$. It can be observed that there is a greater deviation from the ideal behavior for $r=15$ \AA{} and as the distance between the molecules starts to increase the deviations decreases. From the plot of the relative error we can observed that the behavior is not equal when the angles are supplementary, this is due to the fact that the molecules are not symmetric.

When the distance between the pigments is $15$ or $20$ \AA{} the largest percentage of deviation from dipolarity is when the pigments are aligned and the lower deviation is when the acceptor molecule is oriented at $\pi/4$ rad, this is because this orientation allows closer approximation between the pigments. When the distance between the molecules is $30$ and $50$ \AA{} the deviation from the dipolarity begins to decrease and when they are aligned the deviation is of $5$\%. Once again, it is worth noting that this is not a minor result, since most of the photosynthetic pigments found in nature (Fig.~\ref{fig4}) are located $30$ \AA{} from each other, in some antenna complexes the distances are even lower. Also, most of the models used to study the EET in these systems are based in the exciton Hamiltonian, where the electronic coupling between the pigments is essentially dipolar \cite{Aspuru-Guzik2008, Schulten2011,Ai2013}, therefore, if the dipolar interaction is used to describe the interaction between the chromophores the associated error will of the order of $15$\% for chromophores at distances between 20 and 30\AA. Finally, it can be noted that when the orientation of the acceptor molecule is about $\pi/6$ rad the deviation is bigger than when the two chromophores are aligned. This deviation from the dipolar interaction is due to systematic errors made when the oscillations amplitudes of the dipole moment of the acceptor molecule is fitted to a quadratic function. The error is greater in this orientation because the electronic coupling and the magnitude of the dipole moment of the acceptor molecule are lower than those obtained for the collinear orientation.

\section{Conclusions}

In this work we describe the energy transfer between two Chla molecules in real time on the basis of a fully quantum time dependent model as a function of of their distance and orientation. Results for the time dependence of the dipole moment of the two components of the dimer are obtained when one of the pigments is illuminated with a time dependent field in tune with its $Q_y$ excitation. Results for the time dependent variation of the dipole moments of the illuminated and non-illuminated monomer are interpreted in terms of a simple two level system scheme. The growth of the dipole signal with time that results from the full dynamical evolution of the dimer density matrix within the DFTB model is the compared to that expected from the consideration of a dipole-dipole coupling between pigments. Our results indicate that significant differences exist between the simple dipole description an the DFTB Hamiltonian which are more important for dimers placed at distances of about $30$ \AA{} or closer, which are not uncommon in antenna complexes found in nature. 

\section{Acknowledgements}

We acknowledge support by Consejo Nacional de Investigaciones Cient\'{\i}ficas y T\'ecnicas (CONICET) through grant PIP 112-200801-000983. All calculations where performed with a generous time allocation on the supercomputers at the Centre for High Performance Computing at the National University of C\'ordoba. M. B. O. is grateful for a studentship from CONICET. We thank Prof. Marcus Elstner for the provision of DFTB+ Slater-Koster tables for magnesium containing molecules from reference \cite{Cai2007LRCElstner} and the corresponding reference densities.

\bibliography{paper}

\end{document}